%% file: paper.tex
\documentclass[conference,compsoc]{IEEEtran}

\usepackage{url}

\usepackage[breaklinks]{hyperref}
\usepackage{breakurl}
\usepackage{array}
\usepackage{color}
\usepackage{cite}
\usepackage{pifont}
\usepackage{paralist}
\usepackage{float}
\usepackage{tikz}
\usepackage{tikz-qtree}
\usepackage{amsfonts}
\usepackage{amsmath}
\usepackage{amssymb}
\usepackage{mdwlist}
\usepackage[ruled,vlined]{algorithm2e}
\usepackage{enumitem}
\usepackage[all]{nowidow}
\usepackage{booktabs}
\usepackage{mdframed}
\usepackage{flushend}

\newcommand{\myparagraph}[1]{\noindent{\bf#1.}}

\newcommand{\name}{PKISN\xspace}
\newcommand{\TBTF}{too-big-to-be-revoked\xspace}

\begin{document}

\author{
    \IEEEauthorblockN{Pawel Szalachowski, Laurent Chuat, and Adrian Perrig}
    \IEEEauthorblockA{Department of Computer Science\\ETH Zurich, Switzerland}
}

\title{PKI Safety Net (PKISN):\\ Addressing the Too-Big-to-Be-Revoked Problem of
the TLS Ecosystem}

\maketitle

\begin{abstract}
In a public-key infrastructure (PKI), clients must have an efficient and secure
way to determine whether a certificate was revoked (by an entity considered as
legitimate to do so), while preserving user privacy. A few certification
authorities~(CAs) are currently responsible for the issuance of the large majority of
TLS certificates. These certificates are considered valid only if the
certificate of the issuing CA is also valid. The certificates of these important
CAs are effectively \emph{too big to be revoked}, as revoking them would result
in massive collateral damage. To solve this problem, we redesign the current
revocation system with a novel approach that we call PKI Safety Net (PKISN),
which uses publicly accessible logs to store certificates (in the spirit of
Certificate Transparency) and revocations. The proposed system extends existing
mechanisms, which enables simple deployment. Moreover, we present a complete
implementation and evaluation of our scheme.
\end{abstract}

\section{Introduction}
\label{sec:intro}
\input{intro}

\section{Background}
\label{sec:pre}
\input{pre}

\section{\name Overview}
\label{sec:overview}
\input{overview}

\section{\name Details}
\label{sec:details}
\input{details}

\section{Deployment}
\label{sec:deployment}
\input{deployment}

\section{Security Analysis}
\label{sec:analysis}
\input{analysis}

\section{Realization in Practice}
\label{sec:implementation}
\input{implementation}

\section{Evaluation}
\label{sec:eval}
\input{eval}

\section{Discussion}
\label{sec:discussion}
\input{discussion}

\section{Conclusion}
\label{sec:conclusions}
\input{conclusions}

\section*{Acknowledgments}
We thank our shepherd Bart Preneel, the anonymous reviewers, and Franz Saller for their
valuable feedback.
We gratefully acknowledge support from ETH Zurich and from the Zurich
Information Security and Privacy Center (ZISC).

\bibliographystyle{abbrv}
\bibliography{references}

\end{document}

%% file: intro.tex
The TLS public-key infrastructure (PKI) is an essential component of today's Internet,
as it enables to use and verify certificates for secure communications.
Certification authorities~(CAs) are trusted third parties responsible for
issuing and signing digital certificates, which contain authenticated public keys.
To do so, CAs also own certificates. Naturally, the
problem is that private keys can get compromised. As a
consequence, the ability to revoke any certificate (including a CA certificate)
and verify if a given certificate has been revoked is crucial. This process
should be lightweight, secure, and preserve user privacy. To develop a new
revocation scheme, one should also answer the question of who must be able to
revoke a given certificate. The owner, the issuer, the root CA, intermediate
CAs, or a combination of these?

The current revocation schemes simplify many aspects of the PKI ecosystem. The
most striking example is the following. As any private key can be stolen, it
should be possible to revoke any certificate; however, just like some
corporations are said to be ``too big to fail'', the certificates of some
certification authorities are, in practice, \emph{too big to be revoked}. A
study showed that around 75\% of certificates had been issued by only three
different companies, and one specific GoDaddy CA private key had signed 26\% of
all valid certificates in March
2013~\cite{Durumeric:2013:AHC:2504730.2504755,Arnbak:2014:SCH:2668152.2673311}.
Revoking this particular certificate, if the corresponding private key were
compromised, would mean that 26\% of all websites that use HTTPS would be
unavailable (or accessible only if the security warning displayed by
browsers is ignored). Diginotar and Comodo are two infamous examples of
attacked CAs. After the breach, while Diginotar had its certificate revoked
and removed from most CA lists~\cite{MozillaDigiNotarRemoval} (before eventually
declaring bankruptcy), Comodo's incriminated CA certificate was not revoked and
is still present in CA lists~\cite{MozillaList}.
The Comodo Group is leading the certificate issuance business (with a
market share of 33.9\%, and some single private keys that have been used to sign the
certificates of 5.5\% of all the websites considered in a W3Techs survey from
February 2015~\cite{W3Techs}). Consequently, Comodo's root certificate could not
be revoked without effectively preventing users from establishing TLS
connections with a significant
portion of the Internet. For this reason, we claim that the TLS revocation system needs to be
redesigned to remove the collateral damage that would be induced by the revocation of
some CA certificates.

We observe that time is a key element in this problem, as certificates
issued before the certification authority was compromised should not become
invalid when the CA certificate is revoked. Therefore, we suggest that introducing
a timestamp server for certificates and revocations can prevent invalidating
any legitimate certificate. Another problem of current revocation schemes is
that a certificate cannot be directly revoked by its owner. Our approach,
PKI Safety Net~(PKISN), solves this issue. In PKISN, domain owners can use their
private keys to revoke the corresponding certificates, while CAs can use a dedicated
revocation key, which can be securely stored offline as it is not needed during
normal operation.

The scheme we present is inspired by log-based solutions such as
Certificate Transparency (CT), which was recently introduced and
deployed~\cite{rfc6962}. CT aims to
make the actions of CAs more transparent by introducing a log that makes
certificates publicly visible. However, CT mainly covers the issuance aspect of
the problem. We propose to improve the system in several ways, in particular, by
using two distinct hash trees (the data structure on which CT relies) to store
not only certificates, but also revocations. Moreover, \name gives CAs
the ability to revoke their own certificates after a certain point in time to ensure
that previously-signed certificates remain valid.

Security breach notification laws require CAs to notify relevant
authorities about a data breach. For example, in the E.U., this notification
must occur within 24 hours. In other words, certification authorities are
compelled by law to take rapid action to disclose a data breach when
it is detected. This notification should be immediately followed (if not
preceded) by a set of measures that can mitigate the attack, but it is currently
not possible to simply revoke certain compromised CA certificates without
incurring substantial collateral damage.

The major contributions of this paper are the following.
Through \name, we redesign the revocation system to better express the
hierarchical structure of certificates, rebalance the power of PKI actors, and
address the too-big-to-be-revoked problem.
In addition to the existing deployment plans of CT, we propose and discuss
new models that maximize privacy and allow to monitor the log in a lightweight
manner, and we show how the log can be designed to handle these deployment models.
We present an evaluation and a full implementation of our system.

%% file: pre.tex
\subsection{The TLS Public-Key Infrastructure}
In TLS, certificates form a chain of trust (\textit{certificate chain}) that
starts with the root CA's self-signed certificate and ends with the server's
certificate. This chain can contain a number of intermediate CAs and each
certificate in the chain (except the root) is signed by the private key
corresponding to the public key of the parent certificate. TLS clients
(e.g., browsers) need a list of root CA certificates considered trustworthy
to initiate the verification of other certificates. The \texttt{basicConstraints} extension
indicates whether a certificate is a CA certificate. For convenience, we will
use a simple notation to represent the chain of trust formed by a series of
certificates, as follows:
\begin{equation}
C_a \rightarrow C_b \rightarrow \dots \rightarrow C_c,
\end{equation}
where $C_a$ is a root CA certificate, $C_b$ and possibly other certificates are owned
by intermediate CAs, and $C_c$ is an end-entity (leaf) certificate.

Starting 1 April 2015, certificates must not be issued with a validity period greater than
39 months~\cite{CABForum}. However, this concerns only leaf certificates,
i.e., not CA certificates. In fact, certain root CA certificates are
valid for up to 30 years (e.g., the certificate of \emph{CA Disig Root R1}, present in the
list provided by Mozilla~\cite{MozillaList}, will be valid until July 2042).

\subsection{Desired properties}\label{sec:pre:prop}

Here are the properties that we expect of a satisfactory revocation system:

\begin{description}
\item[Efficiency:] transmission, computation, and storage overheads are
    reasonable and the deployment of the system is cost-effective.

\item[Timeliness:] the \textit{attack window}, i.e., the time between the detection of an
    attack and the moment when the corresponding certificate is considered invalid
    by all clients, is short (ideally, on the order of minutes/hours).

\item[Privacy:] clients can obtain certificate-validity information without sacrificing their privacy.
    In particular, users should not be forced to contact any other party than
    the server they connect to in order to obtain the certificate status.

\item[Authenticity:] only legitimate parties can create a revocation message
    for a certificate, but that message is verifiable by everyone.
    The set of legitimate parties depends on a revocation policy.

\item[Independence:] the revocation is independent from the circumstances in which
    the process takes place (e.g., server configuration or the availability of a
    special third party).
    Ideally, whenever an allowed entity has issued a revocation message,
    and a certain server is accessible, then clients of this server should be able
    to access the disseminated message.
    An adversary must not be able to suppress a revocation.

\item[Complete status information:] revocation messages must provide the status of
    all certificates in the chain of trust.

\item[Transparency:] revocations must be publicly accessible and persistent,
    to guarantee to the interested parties that, when a revocation is
    successfully issued, it is impossible to claim that the certificate is still
    valid.

\item[Backward availability:] the revocation system must solve the
    too-big-to-be-revoked problem of the current TLS PKI. In other words, it must
    be possible to revoke any CA, without causing collateral
    damage, i.e., without revoking certificates that were legitimately issued before
    the CA's private key got compromised.
\end{description}
(The efficiency of \name is evaluated in \S\ref{sec:impl:perf},
\S\ref{sec:eval:storage}, and \S\ref{sec:eval:bw},  while security properties
are discussed throughout \S\ref{sec:analysis}.)

\subsection{The Evolution of Revocation Schemes and their Drawbacks}
\label{sec:pre:related}

\begin{inparaenum}[\itshape a\upshape)]
The first attempt to address the revocation problem was realized with Certificate
Revocation Lists (CRLs)~\cite{ccitt1988509}, published by CAs at CRL
distribution points. To verify the validity of a certificate, the browser downloads a CRL
and checks whether the certificate is listed. Unfortunately, the CRL approach has
many drawbacks:
\item It is inefficient, since the entire CRL must be downloaded to verify a
single certificate (a $n$-certificate chain requires $n$ connections).
\item CAs can violate the privacy of users by creating a dedicated distribution
point for a target certificate. Whenever a user connects to this special
distribution point, it means that this user is very likely to visit the website
that corresponds to the target certificate.
\item Gruschka et al.~\cite{gruschka2014analysis} reported that, during a 3-month
period, only 86.1\% of the CRL distribution points had been available.
Mainly due to efficiency issues, the usefulness of CRLs was
questioned several years ago~\cite{rivest1998can,conf/fc/McDanielR00,iliadis2003towards}.
\end{inparaenum}

There are many schemes that improve the format of standard CRLs.  For instance,
Kocher~\cite{kocher1998certificate} proposed to use a Certificate Revocation Tree. This
data structure, based on binary hash trees, allows to efficiently prove that a
certificate is not revoked. Naor and Nissim~\cite{naor2000certificate}
suggested a similar solution, and their Authenticated Dictionaries support
certificate insertion and deletion more efficiently. Unfortunately, these
methods have not been adopted.

To address the inefficiency of CRLs, the Online Certificate Status Protocol
(OCSP)~\cite{santesson2013x} was proposed.  In OCSP, clients contact a CA to get
the status of a certificate. However, this solution is still inefficient (the CA
may be under heavy load, and an extra connection is required), and has a serious
privacy issue (the CA learns about the server that the browser is contacting). OCSP
Stapling~\cite{pettersen2013transport} solves these problems. In OCSP Stapling,
the server periodically obtains an OCSP response from its CA, and then sends
the response along with the certificate in subsequent TLS connections.
Unfortunately, the deployment and effectiveness of this technique depend on the server
configuration (e.g., the age of a stapled response can be customized by a configuration
parameter, which may introduce a long attack window).
Liu~et~al. reported~\cite{liu-2015-revocation} that only 3\% of certificates are
served by servers supporting OCSP Stapling.
Moreover, OCSP and OCSP Stapling only return the status of a single certificate
(not the entire chain). To address this problem, an
extension~\cite{pettersen2013transport} was proposed.

Recently, browser vendors decided to disseminate special CRLs (called CRLSets)
through software updates~\cite{langley2012revocation,mozilla_rev}. Such an approach
does not require any server reconfiguration,
but CRLSets only support certain \textit{Extended Validation} (EV) certificates~\cite{langley2012revocation}.
Such a policy restricts the deployability and effectiveness of the method, as
the fraction of EV certificates is relatively
small~\cite{Holz:2011:SLT:2068816.2068856,Durumeric:2013:AHC:2504730.2504755},
and the revocation process is still conducted through a CA (a user cannot revoke his own
certificate without contacting the CA).
A study showed that Chrome's CRLSet contains only
0.35\% of all revoked certificates~\cite{liu-2015-revocation}.

Short-Lived Certificates (SLCs)~\cite{rivest1998can,topalovic2012towards} solve
problems associated with CRLs and OCSP, by periodically providing domains with
fresh certificates with a limited validity period. SLCs are designed to be
valid for a few days, and as they are irrevocable, a long attack window exists.
SLCs are intended for leaf certificates, hence intermediate and especially root
certificates cannot benefit from the properties of SLCs. In addition, their
deployment depends on server configuration.

Another recent approach, called RevCast~\cite{schulman2014revcast}, improves
revocation dissemination through unique properties of radio broadcast. RevCast
proposes an architecture where CAs broadcast revocation messages and users with
radio receivers can receive them immediately. RevCast employs a blacklist approach
where the user must possess the entire CRL, and to satisfy this requirement an
additional infrastructure must be provided or users have to continuously listen
to broadcast
transmission. RevCast also requires users to purchase and install
radio receivers.

None of the schemes presented above provides the transparency property. Revocation
Transparency~\cite{laurie2012revocation}, which was proposed as a supplement for
Certificate Transparency (see \S\ref{sec:pre:ct}), was the first attempt to provide
that property. Unfortunately, due to the
introduced data structure, checking whether a certificate is revoked might be
inefficient in practice. Additionally, Revocation Transparency lacks a detailed
description.

Log-based approaches such as AKI~\cite{AKI}, ECT~\cite{CIRT},
ARPKI~\cite{ARPKI}, PoliCert~\cite{PoliCert}, and DTKI~\cite{yu2014dtki} take the transparency of
revocations into consideration. However, AKI, ECT, and ARPKI do not allow
domains to use multiple certificates (which is a common
practice today~\cite{ssl_lab}). In ECT, only the most recent certificate is
considered valid for any given entity. Similarly, in AKI and ARPKI, a certificate
expresses the domain's policy, which must be unique. Consequently, these
systems are designed in such a way that, at a given point in time, there can
exist only one active certificate per domain name. 
To solve this issue PoliCert decouples
policies from certificates. Similarly, DTKI introduces a \textit{master
certificate} and a \textit{mapping server}, which also allow a domain to possess
multiple certificates.
Unfortunately,
all these schemes (including PoliCert and DTKI) simplify the certificate hierarchy by
ignoring intermediate CAs, and consider that certificates are signed directly
by root CAs. Such certificates are unusual in practice, and taking intermediate
CAs into consideration would introduce a significant complexity to the log and protocol designs.
For instance, to return complete status information, a log would need to efficiently
look up all relevant information about a particular certificate chain (without performing
a linear search).
Furthermore, the previous proposals do not handle revocation of CA certificates.

Unfortunately, none of the methods proposed in the literature identifies and
solves the too-big-to-be-revoked problem of the current TLS PKI and would thus
create large collateral damage if a popular CA certificate were revoked.

\subsection{Certificate Transparency}
\label{sec:pre:ct}

The Certificate Transparency (CT)~\cite{rfc6962} project was initiated by Google
and aims at making the issuance of TLS certificates accountable and publicly
visible. In order to achieve this goal, log servers are used to collect
certificates that can be submitted by anyone (clients, servers, CAs).

The CT framework relies on the Merkle tree (also called
hash tree) data structure. In the binary Merkle trees used in CT, leaves are
essentially hashes of
certificates and the other nodes are obtained by hashing the
concatenation of their two children. We can distinguish between two types of Merkle
trees. When new leaves are generated, they can either be appended
to the tree (in chronological order) or the tree can be continuously
sorted (in lexicographical order).
In CT, logs use append-only trees sorted in chronological order, because
it can be efficiently proven (with a number of nodes logarithmically proportional
to the number of entries in the tree) that a
certificate is part of the tree and that a given tree is the extension of
another tree.
Trees that are sorted in lexicographical order, on the other hand,
allow to efficiently show that a certain entry is absent from the tree.

When a certificate is submitted to the log for inclusion, it returns a 
Signed Certificate Timestamp (SCT), which is a promise to incorporate the
certificate to the tree within a fixed time period called the Maximum Merge Delay (MMD).
The SCT must be provided by the TLS server
to its clients at every connection, and the documentation~\cite{rfc6962} of CT
describes three ways to do so:
via OCSP Stapling, via a TLS extension, or via an X.509v3 extension. The last method
is of particular interest as it is CA-driven (i.e., CAs directly embed the
SCT into the certificate at issuance) and does not require servers to be updated,
but it requires that CAs participate.

\subsection{Assumptions}
For our revocation system to be operational, we make the following assumptions:
\begin{itemize}[leftmargin=0.5cm]
\item It is possible to determine when the private key of a CA is misused, in particular,
by monitoring logs or with audits. (This is easier to achieve if certificate
logging is mandatory, which is the case for \name.)
\item CAs can store a special private key offline in a secure manner.
\item Browser software is provided by a single vendor. (This assumption
is introduced for the sake of simplicity and can be easily relaxed.)
\item Browsers have a working software-update mechanism.
\item The log server is highly available (for both read and update operations) to all parties.
\item The different parties are loosely time-synchronized (up to few minutes),
    and
    time is expressed in Unix seconds.
\item The cryptographic primitives used by PKISN are secure.
\item Only one log server exists, but extending PKISN to
multi-log settings is discussed in \S\ref{sec:discussion}.
\end{itemize}

\subsection{Adversary Model}
\begin{inparaenum}[\itshape a\upshape)]
We consider that an adversary can steal a domain's private key to perform a man-in-the-middle
attack or a CA's private key to issue malicious certificates/revocations, but an attacker
cannot access a CA's offline (revocation) key,
and cannot access key(s) used for software update.
The adversary can also contact the log (to fetch or submit data) as any other party.
The adversary's goal can be to:
\item cause collateral damage and make many websites unavailable,
\item violate the revocation policy and convince a client that a revoked certificate
is still valid, or
\item revoke a valid certificate without legitimately owning the appropriate key.
\end{inparaenum}

\subsection{Notation}
Throughout the paper, we use the notation presented in Table~\ref{tab:notation}.

\begin{table}
\caption{Notation.}
\normalsize
\begin{tabular}{@{~}lp{200pt}@{~}}
\toprule
$C_x$ & certificate \\
$R_{C_x}$ & revocation of certificate $C_x$ \\
$t_x$ & timestamp \\
$\textnormal{\textit{sk}}_x$ & secret key associated with the public key
        authenticated by $C_x$ \\
$\textnormal{\textit{rk}}_x$ & revocation key (stored offline) associated with
        a CA certificate $C_x$ \\
$\textnormal{\textit{vk}}$ & key used by the software vendor \\
$k_{\textnormal{\textit{log}}}$ & log key \\
$H(.)$ & cryptographic hash function \\
$\textnormal{\textit{Sig}}_k(m)$ & message $m$ signed with key $k$ \\
$\varnothing$ & \textit{null} value \\
$\|$ & concatenation \\
\bottomrule
\end{tabular}
\label{tab:notation}
\end{table}

%% file: overview.tex
This section gives a high-level picture of the overall system and introduces the
entities involved and basic terminology.
In \name, clients/browsers want to communicate securely with servers/domains.
A server is authenticated through a certificate chain created
by a number of CAs. All certificates and revocations must be logged by a log server. At every
\textit{update time}, each log updates its local database, and the time period between
these updates is called the \textit{scheduling period}. The log is
verified by browsers and dedicated parties called \textit{monitors}. CAs must also act as
monitors to verify that no illegitimate certificate (issued on their
behalf) is present in the log.

\subsection{The Certificate Log as a Timestamping Service}
The main goal of our work is to solve the \TBTF problem of the current TLS PKI.
Namely, we want to enable revocation of CA certificates without causing
collateral damage. The typical scenario in which a revocation is required is
after a private key compromise. Currently, revocation of a compromised private
key owned by an important CA, should invalidate all certificates signed by this
key, as the certificates may have been fraudulently created.

Our main observation is that when a compromised CA can determine the time of the
attack---more precisely, the time at which an illegitimate action (like
certificate issuance or revocation) was first
observed---then the certificates signed before the attack can still be considered valid.
Only certificates issued after the attack are potentially malicious, and should not be
trusted. It is possible for CAs to determine the time of that attack as, in the PKISN
framework, all certificates and revocations must be logged before they are
considered valid.
Thus, the instant of the first maliciously registered certificate is the instant
of compromise.

As we cannot rely on the creation-time field of a certificate (because it may
be easily predated by the adversary), the main challenge in resolving the \TBTF issue
is the lack of a trusted timestamping service~\cite{Haber91howto}. \name leverages the
concept of a certificate log to provide this service. Depending on the deployment scenario,
a domain or a CA submits the certificate to a log. When the certificate is accepted, the log
returns a \textit{chain commitment}~(CC) and appends the certificate to the
tree in the next update. Additionally, all intermediate certificates are
added as well (if they are not already in the log).  The returned commitment
includes a list of \textit{registration timestamps} that specifies when the
non-registered
certificates in the chain will be present in the log (in this case, the timestamp
denotes the next update time) and when the already registered certificates were appended to
the log.

Every new certificate is appended along with its registration timestamp. Thereafter,
anyone with the obtained commitment can query the log for the presence proof of the certificate.
As a presence proof includes a registration timestamp, it is the confirmation that the log
contained a given certificate at a given point in time. Hence, a requester can
assert that the certificate was created before this timestamp.

\subsection{Transparent and Persistent Revocation}
\name also employs a public log for storing revocations. In order to
enhance the transparency of the current PKI ecosystem, revocations
need to be logged. For instance, whenever a key is compromised or lost, the
owner should have a guarantee that a revocation will be visible for others at
least until the revoked certificate expires. The obligation of logging
certificates also makes CAs more transparent, as they cannot misbehave by
distributing two different CRLs or two different OCSP responses.

\name introduces special types of revocation messages (see
\S\ref{sec:rev_msg}), and due to the hierarchical nature of the certificate
chain, a given certificate can be revoked by a set of entities (see \name's revocation
policy below).

An authorized entity (usually the owner of a certificate) who wishes to revoke a
certificate can create a special revocation message. This message is submitted to
the log, which, after verification, returns a \textit{revocation commitment}
stating that the revocation will be appended to the log in the next update. When the
revocation message is in the log, the presence proof for the corresponding
certificate must contain this revocation message. To minimize the
attack window, whenever a revocation is pending for addition, the log can
accompany the presence proof of a certificate with its revocation,
without waiting for the end of a scheduling period.

\subsection{Revocation Policy}\label{subsec:revocation_policy}
In the current PKI ecosystem, a certificate can be revoked only by two
parties, namely the issuer and a software vendor (e.g., a browser or an
OS vendor). 
Whenever a domain wishes to revoke its own certificate, the domain must contact the
appropriate CA that will eventually issue the revocation. Obviously, such a procedure
results in a prolonged attack window and depends completely on the issuing CA. Alternatively,
the software vendor can simply blacklist certain certificates and
propagate the changes through software updates. In this case,
a domain has to contact the software vendor. This option can be also used for revoking
misbehaving CAs. However, software vendors are reluctant to use this option, as
it renders all servers with a certificate issued by that particular CA unavailable.

\name introduces a revocation policy that reflects the interactions of the current PKI
and the hierarchical structure of the certificate chain. Specifically, we introduce the
following revocation rules:
\begin{description}
    \item[The owner] of a leaf certificate can revoke this certificate using the associated
        private key.\footnote{By \textit{associated private key} we mean the one
        corresponding to the public key that the certificate authenticates.}
        This option gives domains the opportunity to revoke, without the need
        to contact CAs or a software vendor.

    \item[The issuer] (or an upper-level issuer, i.e., a CA in the certificate chain)
        of a leaf certificate can revoke that certificate. The revocation message is created by the
        issuer's (i.e., a CA's) private key and can be performed, for
        example, when a domain lost its private key. Note that a certificate can be
        revoked directly by a root CA, without involving intermediate CAs.

    \item[CAs] can revoke their own certificates and the
        certificates of their child CAs from a given point in time called
        \textit{revocation timestamp}. This revocation states that certificates
        and revocations issued after a revocation timestamp should be considered
        invalid and should be ignored during the certificate-chain validation.
        The CA's own certificates are revoked with a dedicated \textit{revocation key}, while
        child certificates are revoked with a regular private key.
        With the revocation key, a CA can prevent all
        potentially malicious actions starting from a certain point in time.
        A CA can use its revocation key only once, and as it invalidates the
        CA's certificate there is no need to revoke or update a revocation
        key. A new revocation key is generated every time a CA's certificate is
        created.

    \item[A software vendor] can revoke any certificate, and for CA
        certificates, they have to specify a revocation timestamp as above. Only
        child certificates and revocations issued with the revoked
        certificate before that timestamp are considered valid.
        Currently, software vendors effectively have the ability to revoke
        any certificate, so this option explicitly reflects their power in the current
        TLS PKI ecosystem. Moreover, \name holds their actions accountable and
        transparent. For the sake of simplicity, we assume that there is a
        single software vendor that issues revocations with a private \emph{vendor key}.
        The corresponding public key is provided to the clients within the software
        (like today), and can be updated with a software update (but cannot be
        revoked through \name).
\end{description}
Note that we do not allow a revoked (i.e., compromised, usually) CA to revoke its child certificates,
even if the revocation had been legitimate 
(otherwise an adversary could cause collateral damage by invalidating
certificates with the already revoked key).
In such a case, any non-revoked
CA in the certificate chain can still issue a valid revocation for the leaf certificate. 
However, after a CA is revoked, its clients should be informed that, although
their legitimately-issued certificates are still valid and can be used, the CA lost its revocation
ability, and the certificates should be reissued in the near future (e.g., few days or weeks).

\begin{figure}[h]
  \centering
  \includegraphics[width=\linewidth]{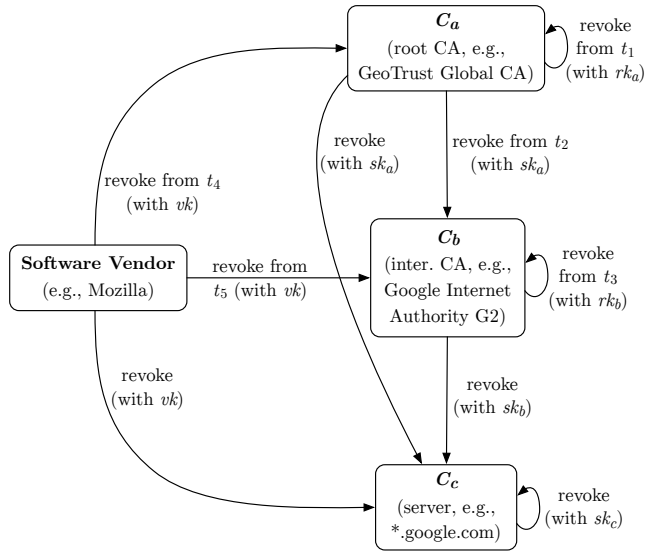}
  \caption{All possible revocations for a certificate chain $C_a\rightarrow
      C_b\rightarrow C_c$, where $C_a$ and $C_b$ have associated keys $\textit{sk}_a,
      \textit{rk}_a$ and $\textit{sk}_b, \textit{rk}_b$, respectively (standard and revocation private keys),
      while the leaf certificate is associated only with a standard private key $\textit{sk}_c$,
  and $\textit{vk}$ denotes the software vendor key.}
  \label{fig:revocations}
\end{figure}

Possible revocation actions for an example certificate chain are presented
in Fig.~\ref{fig:revocations}.
A single certificate can have many associated revocations. All these revocations
can be fetched from the log with a presence proof.

\subsection{Validation}
For a successful validation, a client must be provided with a certificate chain,
a corresponding chain commitment (CC), and a proof from the log. 
First, the input data is pre-validated (for details see \S\ref{sec:pre_valid}),
and verified. This includes a standard chain validation as executed in modern
browsers. However, the \name validation process goes further, as it 
determines time periods for which CAs were behaving legitimately.

As shown in Fig.~\ref{fig:revocations}, a single certificate can be revoked
by different entities and through different revocation messages. Hence,
to achieve an unambiguous validation of revocation messages, priorities
must be established. \name introduces the following priorities for revocation messages,
from the highest priority to the lowest:
\begin{enumerate}
\item revocations issued by the software vendor,
\item revocations created with a dedicated revocation key (only applies to non-leaf certificates),
\item revocations issued by parent CAs,
\item revocations created with the standard private key associated with the
certificate (only applies to leaf certificates).
\end{enumerate}

To conduct a validation, \name introduces the notion of a \textit{legitimacy
period}, which denotes a time period during which actions performed by
CAs are considered valid. The legitimacy period is 
defined between the moment when a certificate is received by the log for the
first time (registration timestamp) and the moment when it expires or is legitimately
revoked (revocation timestamp). A certificate is considered valid when it
passes the \textit{pre-validation} and when all certificates in the chain
were issued (and never revoked) during corresponding legitimacy periods.

\begin{figure}[h!]
  \centering
  \includegraphics[width=\linewidth]{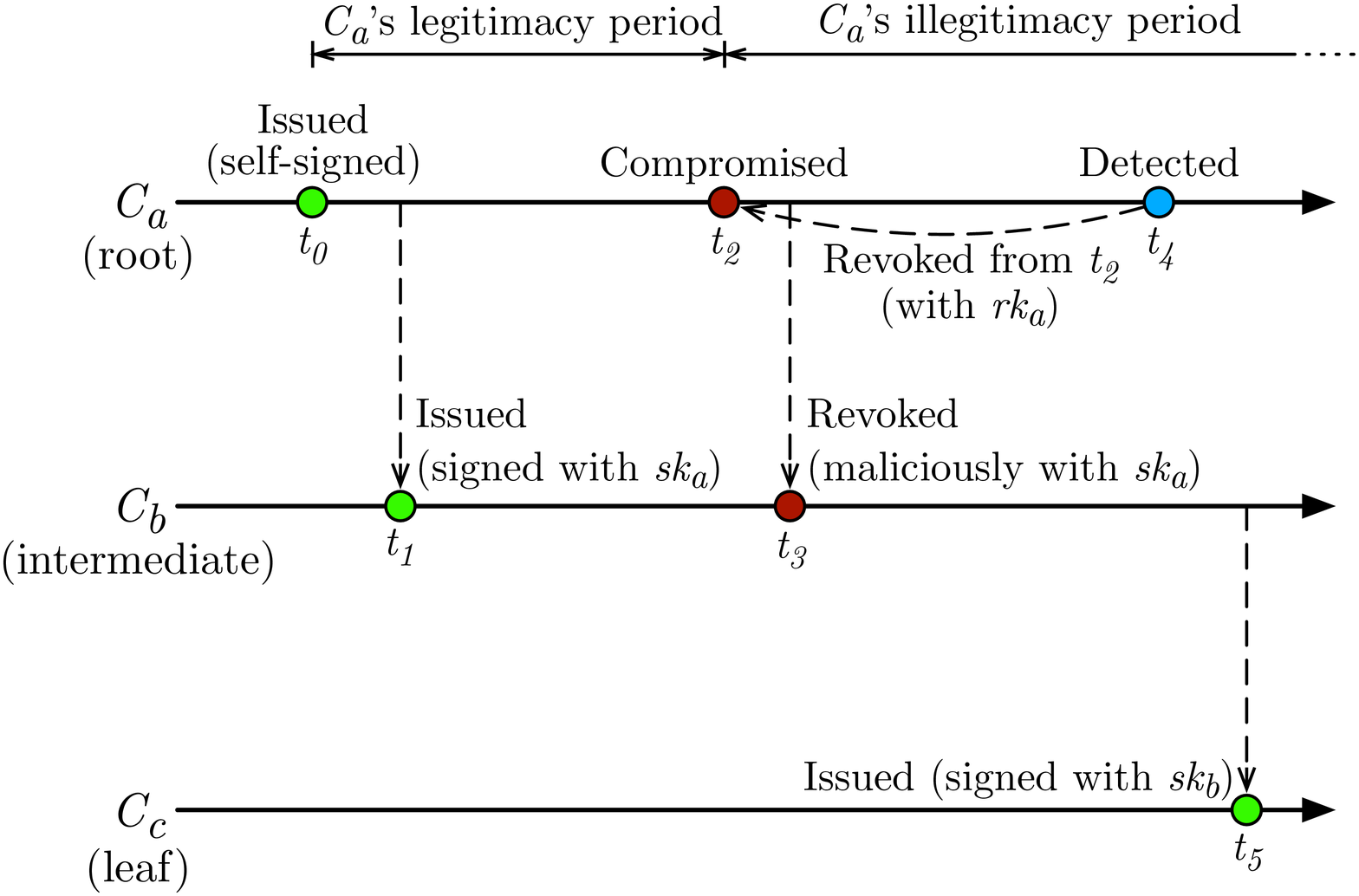}
  \caption{Timelines for a chain of three certificates, with an attack (against the root CA) and a detection thereof.}
  \label{fig:timelines}
\end{figure}

An example that illustrates the concept of legitimacy periods is presented in
Fig.~\ref{fig:timelines}. In this example, the root CA certificate $C_a$ gets compromised,
but the attack is then detected and the CA is able to determine the time at which the attack
was performed. In the meantime, the adversary used the private key to maliciously revoke
the certificate $C_b$ of an intermediate CA.\footnote{Although such an attack was never observed
in the real world (to the best of our knowledge), nothing currently prevents an adversary who
compromised a private key from performing revocations. Therefore, our new scheme should
take this case into account.}
In this particular case, the leaf certificate $C_c$ is
valid even though its parent CA certificate was revoked, as \name allows to express the fact
that $C_b$ was maliciously revoked (the revocation was done during the
\emph{illegitimacy period} of $C_a$).

\subsection{Log Consistency}
Periodically, a browser contacts a random monitor to ensure that they share the
same view of the log. As monitors have a copy of the log, they can inform about
historic versions. To prevent equivocation, browsers can compare log information
obtained during the TLS connections with corresponding monitor statements. Even
if such
a procedure does not completely protect against malicious logs, it enables to
detect log misbehavior.

%% file: details.tex
\subsection{Revocation Messages}\label{sec:rev_msg}
\name introduces a new dedicated revocation
key pair for CAs.
The revocation private key is only used when a given CA notices that its
standard private key (used in production) has been compromised or lost.
As the revocation key is not used in production, it should be securely
stored offline.

\name supports two formats of revocation messages. The first one is used
for invalidating leaf certificates:
\begin{equation}\label{eq:rev1}
    R_{C_x} = \textit{Sig}_k(H(C_x), \texttt{revoke}),
\end{equation}
where $k$ can be:
\begin{inparaenum}[\itshape a\upshape)]
\item a private key associated with the authenticated public key in a leaf certificate,
\item the private key of one of the CAs in the certification chain, or
\item a software vendor key. Note that in contrast with the current
revocation system, domains can revoke their own certificates without any
interaction with the issuing CAs. Leaf certificates do not contain a special
revocation key and can be revoked without a revocation timestamp, as they cannot
cause collateral damage.
\end{inparaenum}

Because CA certificates introduce collateral damage,
they are always revoked by the following revocation message:
\begin{equation}\label{eq:rev2}
    R_{C_x} = \textit{Sig}_k(H(C_x), \texttt{revoke from } \textit{rev\_timestamp}),
\end{equation}
where $k$ can be
\begin{inparaenum}[\itshape a\upshape)]
\item the CA's revocation key,
\item the standard private key of a parent CA, or
\item a software vendor key.  This revocation message contains a
revocation timestamp, that indicates a time from which all actions
(certificates and revocations issuances) of the revoked CA must be considered
invalid. This timestamp must be earlier than the expiration time specified within the
revoked certificate.
\end{inparaenum}

\subsection{Structure of the Log}
In \name, a log stores all the issued certificates and revocations, and additionally
processes them like a timestamping service. 
On demand, the log can produce efficient
proofs about the stored content. The log is designed to support the following
operations:
\begin{enumerate}
    \item prove that a given certificate or revocation is in the log and was
        appended to the log at a given point in time,
    \item prove that a given certificate or revocation was not appended to the
        log at a given point in time,
    \item with a given chain commitment (CC), prove that all certificates from
        the chain were appended correctly (according to the timestamps of the CC)
        and show all revocations associated with these certificates, 
    \item prove that one snapshot of the log is an append-only extension of any
        previous one.
\end{enumerate}

\begin{figure*}[t!]
\centering
  \begin{tikzpicture}[scale=.45,every tree node/.style={font=\LARGE,anchor=base}]
    \Tree 
    [.\textit{root}
        [.\fbox{$H_{abcdefgr_0}$}
     [.${H_\mathit{abcd}}$
        [.${H_{ab}}$
        [.$H_a$ $C_{a},t_0$ ] [.$H_b$ $C_{b},t_0$  ] ]
        [.${H_{cd}}$
        [.$H_c$ $C_{c},t_0$ ] [.$H_d$ $C_d,t_0$ ] ] 
    ]
    [.${H_\mathit{efgr_0}}$
        [.${H_{ef}}$
        [.${H_e}$ $C_e,t_0$ ] [.$H_f$ $C_f,t_0$  ] ]
        [.${H_\mathit{gr_0}}$
        [.$H_g$ $C_g,t_0$ ] [.$H_{r_0}$ $r_0,t_0$ ] ] 
    ]
 ]
  [.$H_\mathit{hijd'klmr_1}$
    [.\fbox{$H_{hijd'}$}
        [.${H_{hi}}$
        [.${H_h}$ $C_h,t_1$ ] [.$H_i$ $C_i,t_1$  ] ]
        [.${H_{jd'}}$
        [.$H_j$ $C_j,t_1$ ] [.$H_{d'}$ $R_{C_d},t_1$ ] ] 
    ]
    [.${H_\mathit{klmr_1}}$
        [.\fbox{$H_{kl}$}
        [.${H_k}$ $C_k,t_1$ ] [.$H_l$ $C_l,t_1$  ] ]
        [.${H_{mr_1}}$
            [.\fbox{$H_m$} $C_m,t_1$ ] [.$H_{r_1}$
        \node(r1t){$r_1,$\fbox{$t_1$}}; ] ] 
    ]
  ]
 ]
 \begin{scope}[shift={(20,-1.5cm)}]
    \Tree 
    [.\node(r1f){$r_1$};
        [.$H_{12}$
        [.\fbox{$H_1$} [.$H_b,\varnothing$ ... ] ] [.$H_2$ [.\fbox{$H_a$,$\varnothing$} $r_a$ ]  ] ]
        [.{} [.\fbox{$H_3$} 
        [.$H_c,\varnothing$ ... ] ] ]
    ]
\end{scope}
 \begin{scope}[shift={(19.5,-5.8cm)}]
     \Tree   
     [.{}
         [.$H^a_{12}$
             [.\fbox{$H^a_1$} [.$H_e,\varnothing$ ... ] ] [.$H^a_2$
                 [.\fbox{$H_d,(R_{C_d},t_1)$} $r_d$ ]  
        ] ]
        [.{} [.\fbox{$H^a_{3}$} [.$H_f,\varnothing$ ... ]
         ]  
        ]
    ]
\end{scope}
\begin{scope}[shift={(18.75,-10.2cm)}]
     \Tree   
     [.{}
         [.$H^d_{12}$
             [.\fbox{$H^d_1$} [.$H_k,\varnothing$ $\varnothing$ ] ] [.$H^d_2$
                 [.\fbox{$H_{m},\varnothing$} \fbox{$\varnothing$} ]  
        ] ]
        [.\fbox{$H^d_{34}$}
             [.$H^d_3$ [.$H_g,\varnothing$ $\varnothing$ ] ] [.$H^d_4$
                 [.$H_j,\varnothing$ $\varnothing$ ]
        ] ]  
    ]
\end{scope}
\draw[->](r1f) .. controls +(north:5) and +(south:5) .. (r1t);
\begin{scope}[dashed]
    \draw[->](-10,-6) -- node[below] {\textit{time}} (10,-6);
\end{scope}
\node [draw] at (0,-11) {%
    \begin{minipage}{0.50\textwidth}
        Presence proof for $C_m$ (in the chain $C_a\rightarrow C_d\rightarrow C_m$): 
\begin{multline}\label{eq:proof}
    \{H_m,\varnothing,\varnothing, H^d_1, H^d_{34}, H_d, (R_{C_d},t_1), H^a_1, H^a_3, \\
        H_a,\varnothing, H_1, H_3, t_1, H_m, H_\mathit{kl}, H_\mathit{hijd'},
    H_\mathit{abcdefgr_0}\}.
\end{multline}
    \end{minipage}
};
\end{tikzpicture}
\caption{Example of log trees. The TimeTree stores all objects in chronological
    order, while the leaves of the RevTree's subtrees are sorted lexicographically.
    The log contains one revocation message $R_{C_d}$ associated with the
    certificate $C_d$. Nodes in boxes are needed for the presence proof of
    certificate chain $C_a\rightarrow C_d\rightarrow C_m$.}
\label{fig:trees}
\end{figure*}
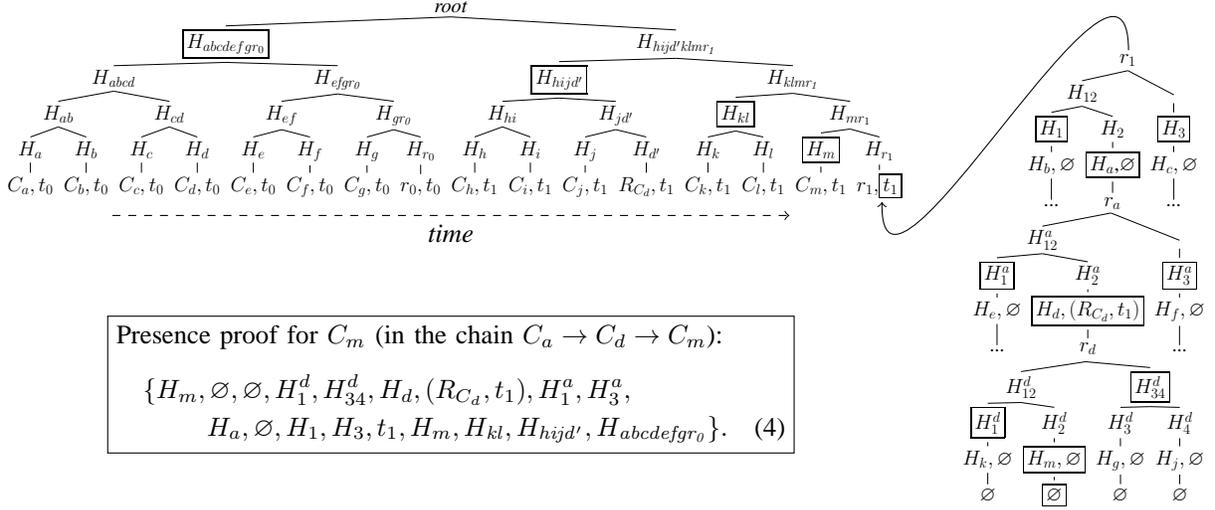

Additionally, relevant information about a particular certificate chain must
be processed efficiently. To provide these features, the log maintains two
hash-tree-based data structures: a \textit{TimeTree} and a \textit{RevTree}
(Revocation Tree). Fig.~\ref{fig:trees} depicts an example of these trees.

The TimeTree contains all objects added to the log in chronological
order. It stores certificates ($C_x$), revocations ($R_{C_x}$), and roots of the
RevTree. All objects are accompanied with a registration timestamp that denotes
when the object was actually appended to the tree. 
With the TimeTree, it is possible to prove that a given object is indeed an
element of the log and was inserted at a given registration timestamp.
Additionally, it is possible to prove that one version of the TimeTree is an
extension of the previous TimeTree.

The RevTree consists of sub-trees (it is, in fact, a forest) that reflect the
hierarchical structure of certificate chains. The RevTree is built
after every scheduling period, and the root of this tree is the last element appended to the
TimeTree in every update of the log. The leaves of every sub-tree consist of:
\begin{itemize}
    \item A hash $H_x=H(C_x\|t_x)$ that identifies a certificate $C_x$ and
        a registration timestamp $t_x$. The leaves of every sub-tree are sorted
        in lexicographical order of these hashes.
    \item The possible revocation messages of $C_x$. This may be
        $\varnothing$ when a certificate has no associated revocation message. Every
        revocation $R_{C_x}$ is accompanied with a registration timestamp, which indicates
        when the revocation was appended to the log (this is not a
        revocation timestamp as in Eq.~(\ref{eq:rev2})).
    \item The root ($r_x$) of the sub-tree (the sub-tree contains certificates
        signed with the private key associated to $C_x$), that may be $\varnothing$ when
        the certificate does not have any children (e.g., leaf certificates). For
        efficiency reasons, leaves can also store pointers to their sub-trees. 
\end{itemize}
The RevTree's top sub-tree identifies root certificates, and every leaf is
associated with a sub-tree of its child certificates. This design allows
the log to efficiently:
\begin{inparaenum}[\itshape a\upshape)]
\item prove that all certificates from a chain were appended to the log at a given
time, and
\item show all the revocations associated with these certificates.
As all the leaves of a RevTree's sub-trees are sorted in lexicographical order, it
is also possible to prove that a given certificate was not appended to the tree at
a given time. In combination with a TimeTree, a complete proof contains the
information that the RevTree's proof comes from the current version of the
RevTree, as its root is the very last element of the TimeTree.
\end{inparaenum}

\subsection{Interactions with the Log}
\myparagraph{Certificate Registration}
Before a certificate is used it must be submitted to the log.
For instance, a domain with leaf certificate $C_m$ sends the following certificate
chain to the log:
\begin{equation}\label{eq:chain}
C_a\rightarrow C_d\rightarrow C_m.
\end{equation}
To automate this operation, certificates can also be submitted to the log by CAs,
in a similar way as \emph{pre-certificates} can be submitted in CT~\cite{rfc6962}.
The log verifies the chain, and schedules the inclusion of non-appended
certificates from the chain to the TimeTree and the RevTree.
Any new certificate will be appended along with a
registration timestamp. The log
returns a \emph{Chain Commitment}~(CC) signed with $k_{\textit{log}}$ immediately after verification.
The CC consists of:
\begin{equation}\label{eq:CC}
    \textit{Sig}_{k_{\textit{log}}}(H(C_m), t_m, t_d, t_a).
\end{equation}
It constitutes a promise that $C_m$ will be appended at $t_m$, and that $C_d, C_a$ are
or will be visible after $t_d$ and $t_a$, respectively. As each certificate
is unique within a log, the registration timestamps for CA
certificates will often be from the past (as it is likely that these
certificates have been submitted before). The following must always be
satisfied: $t_m\ge t_d\ge t_a$.

\myparagraph{Proof Querying}
In the first update time after a successful submission, the certificate is added to the
trees. Thereafter, anyone can query the log for the presence proof of the certificate.
The internal design of \name optimizes the log for serving presence/absence
proofs to a
requester with a certificate chain and a corresponding CC, as this is the most
common interaction with the log.

For instance, with a certificate chain $C_a\rightarrow C_d\rightarrow C_m$ and
the corresponding CC from Eq.~(\ref{eq:CC}), the following request is prepared
and sent to the log:
\begin{equation}\label{eq:req}
    H(C_m\|t_m), H(C_d\|t_d), H(C_a\|t_a).
\end{equation}
Due to the structure of the request, the log can efficiently locate the requested
leaves in the RevTree, and generate a presence proof, by showing all
intermediate nodes necessary to build the tree root. For instance, for the
content presented in Fig.~\ref{fig:trees} and the previous request, the log can
produce the presence proof from Eq.~(\ref{eq:proof}) (see
Fig.~\ref{fig:trees}).
Note that whenever a certificate from the chain has some associated revocation messages,
these messages must be contained within the proof.
It guarantees that a revocation status for every certificate from the chain is
known.
The proof is returned to the requester accompanied with the current signed root:
\begin{equation}\label{eq:root}
    \textit{Sig}_{k_{\textit{log}}}(root, t_x).
\end{equation}
The signed root can also be requested separately. In our setting, the combination
of a presence proof and the signed root is the most important piece of information
from a client's perspective; it contains almost everything to perform a certificate
validation. However, in \name, a log is also obligated to provide extension proofs
between two versions of the TimeTree (to prove the consistency of two snapshots of
the log).

\myparagraph{Certificate Revocation}
An entity allowed to revoke a certificate ${C_x}$ can create a revocation message
from Eq.~(\ref{eq:rev1}) or Eq.~(\ref{eq:rev2}). The revocation message
$R_{C_x}$ is sent along with a certificate chain whose last certificate is
intended to be revoked. The log, after verifying whether the revocation is legitimate
and matches the certificate, schedules the revocation and returns a message:
\begin{equation}\label{eq:rev_commit}
    \textit{Sig}_{{k}_{\textit{log}}}(H(R_{C_x}), t_x),
\end{equation}
which states that the revocation will be appended to the log after time $t_x$.
However, during the scheduling period (when the revocation is not yet
appended) the log can attach a revocation message to every relevant presence
proof. This would reduce the attack window. During the update of the log,
the revocation message is appended to the TimeTree and is appended to the 
RevTree's leaf which corresponds to the revoked certificate. From that time
forward, every presence proof requested for a chain that contains the revoked
certificate must contain the revocation message.

\myparagraph{Monitoring}\label{sec:monitoring}
The role of a \emph{monitor} is to verify the correct behavior of a log. Each monitor
periodically (after every log update) contacts the log and downloads the newly
appended objects and the current signed root. Then, the monitor updates its own
copy of the log, by appending new certificates and revocations to the TimeTree,
and by introducing all changes to the RevTree. After that, the monitor puts the current
root of its RevTree as the last leaf into the TimeTree. Finally, the monitor computes
the root of its own copy of the TimeTree and compares it with the root received from
the log. During this update, the monitor also verifies whether the certificates
and revocations accepted by the log were legitimate. 

Through this periodic update, the monitors can detect any inconsistency/misbehavior
of the log. Anyone can request signed roots from a monitor, and report a proof of
misbehavior such as:
\begin{itemize}
    \item an incorrect CC (with incorrect registration timestamps or absence proof of a
        certificate that was not appended), 
    \item a revocation that is not appended (showing a message from
        Eq.~(\ref{eq:rev_commit}), and a proof that the revocation is not in the
        log),
    \item two different roots from the same time period,
    \item the presence proof of an invalid certificate or revocation. 
\end{itemize}

The monitor, in such a setting, must replicate the log's content. In
\S\ref{sec:light_monitor}, we propose a novel deployment model that allows to
implement a monitor in a lightweight manner. 

\subsection{Validation}\label{sec:validation}
To conduct a certificate validation, a client needs:
\begin{inparaenum}[\itshape a\upshape)]
    \item a certificate chain, 
    \item a chain commitment, 
    \item a proof of presence, 
    \item and the corresponding signed root. 
\end{inparaenum}
The full validation is presented in Algorithm~\ref{alg:validation}. This section
presents the different steps. We assume that before validation, the structure and format of
all messages is checked.
\begin{algorithm}[h!]
	\caption{Complete certificate validation.}
	\label{alg:validation}
	\small
	
	\begin{description}
        \item[$\textnormal{\textit{root}}:$] signed root (TimeTree), e.g.,
            Eq.~(\ref{eq:root})
		\item[$\textnormal{\textit{proof}}:$] presence proof, e.g.,
            Eq.~(\ref{eq:proof})
		\item[$\textnormal{\textit{chain}}:$] certificate chain, e.g.,
            Eq.~(\ref{eq:chain})
		\item[$\textnormal{\textit{CC}}:$] signed chain commitment, e.g.,
            Eq.~(\ref{eq:CC})
		\item[$\textnormal{\textit{name}}:$] name of the contacted domain
		\item[$t_x:$] registration timestamp of $C_x$
		\item[$\textnormal{\textit{LP}}:$] dictionary that maps certificates to
            their legitimacy periods
		\item[$\textnormal{\textit{currTime()}}:$] returns current time in Unix seconds
		\item[$\textnormal{\textit{preValidate()}}:$] returns \textit{true} $\Leftrightarrow$ pre-validation passes
		\item[$\textnormal{\textit{verifyProofs()}}:$] returns \textit{true} $\Leftrightarrow$ proof is correct
		\item[$\textnormal{\textit{determineLP()}}:$] returns legitimacy period
            of a certificate 
	\end{description}
	
    \SetKwProg{func}{function}{}{}
    \func{isValid(root, proof, chain, CC, name)}{
    	\If{\textbf{not} preValidate(chain, name)}{
    		\Return{FAIL}\;
    	}
    	\If{\textbf{not} verifyProofs(root, proof, chain, CC)}{
    		\Return{FAIL}\;
    	}
    	\For{$C_x \in \textit{chain}$ \textnormal{/*start from root CA*/}}{
    		$\textit{LP}[C_x] \leftarrow \textit{determineLP}(\textit{LP}, C_x, t_x, R_{C_x}, ...)$\;
    		\If{$C_x$ \textbf{is not} a root certificate}{
    			\If{$t_x \not\in \textit{LP}[C_x.\textit{parent}]$}{
    				\Return{FAIL}\;
    			}
    		}
    		\If{$C_x$ \textbf{is} a leaf certificate \textnormal{/*last certificate*/}}{
    			\uIf{$\textit{currTime}() \in \textit{LP}[C_x]$}{
    				\Return{SUCCESS}\;
    			}
    			\Else{
    				\Return{FAIL}\;
    			}
    		}
    	}
    }
\end{algorithm}

\myparagraph{Pre-Validation}\label{sec:pre_valid}
The first step is to pre-validate the certificate chain against a given domain name.
This is similar to the standard validation procedure executed by modern browsers.
It encompasses checking whether the leaf certificate is issued for the given
domain, checking whether the certificate chain is correct and terminates with a trusted
root certificate. Usually, such a pre-validation also includes expiration
checks, but this functionality is enhanced by \name.

\myparagraph{Proof Verification}\label{sec:proof_verif}
During the next step, the browser verifies the authenticity and correctness of the
obtained log proofs. First, the match between a proof, a certificate chain, and a
chain commitment is verified. The browser checks whether the proof contains (in
correct locations) the hashes of all the chain's certificates concatenated with the
corresponding timestamps (from the CC). Then, by hashing the elements of the proof,
a root is computed and compared with the signed root provided as input.
When the roots are the same, the verification passes, and the signed
root can be kept for further consistency checks and monitoring (\S\ref{sec:monitoring}).

\begin{figure}[b!]
  \centering
  \includegraphics[width=0.84\linewidth]{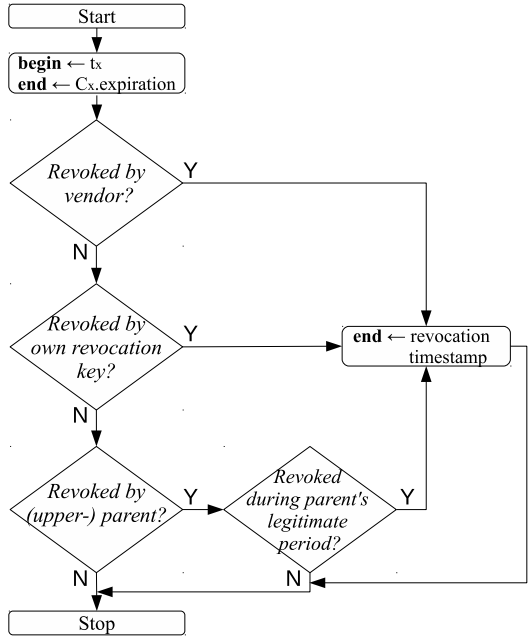}
  \caption{Legitimacy period determination for CA certificates, where $t_x$
  denotes $C_x$'s registration timestamp.  After the algorithm's execution, the
  legitimacy period is expressed as a time range (from \textbf{begin} to
  \textbf{end}).} 
  \label{fig:period}
\end{figure}

\begin{figure}[b!]
  \centering
  \includegraphics[width=0.94\linewidth]{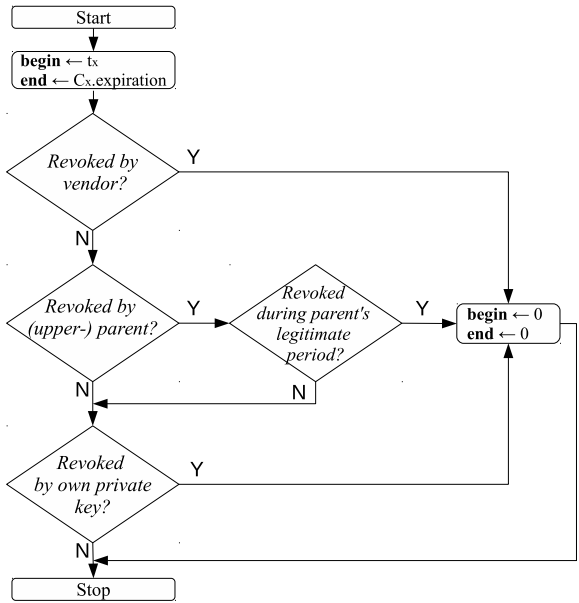}
  \caption{Legitimacy period determination for leaf certificates, where $t_x$
  denotes $C_x$'s registration timestamp.  After the algorithm's execution, the
  legitimacy period is expressed as a time range (from \textbf{begin} to
  \textbf{end}).} 
  \label{fig:period_leaf}
\end{figure}

\myparagraph{Legitimacy-Period Determination}\label{sec:lp_deter}
The next step in the validation procedure is to determine the legitimacy periods of all
certificates in the chain. This procedure slightly differs depending on the type of
certificate (leaf certificates do not introduce any collateral damage and thus are
revoked without specifying a revocation timestamp).
Legitimacy periods are determined as presented in Fig.~\ref{fig:period} (for CA
certificates) and as in Fig.~\ref{fig:period_leaf} (for leaf certificates). 


The procedure starts with the first (the root) certificate in the chain, and is
executed for every subsequent certificate. First, the legitimacy period is set as
a time range from $t_x$ (the registration timestamp) to $C_x.\texttt{NotAfter}$ (which
denotes the expiration time specified within the certificate). If a revocation
issued by the software vendor is present, the legitimacy period of the current certificate
is limited by the time from which the vendor revoked this certificate (i.e., up to the
revocation timestamp). If a certificate is
revoked with a private key associated with the certificate ($\textit{rk}_x$ for a CA certificate and $\textit{sk}_x$ for a leaf certificate), the legitimacy
period is similarly limited by the revocation timestamp from the revocation message. The last option is
a revocation realized by parent CAs. Similarly, the legitimacy period can be
restricted, but this revocation message must be issued during the legitimacy
period of the issuer.

The legitimacy period of a leaf certificate can express two states (revoked or
non-revoked), but the processing logic is similar to the previous case.

During the complete validation procedure (see Algorithm~\ref{alg:validation}), it is
also ensured that every certificate from the chain (except the root) has a
registration timestamp within the legitimacy period of its parent. In the final step of
the validation, it is ensured that the leaf certificate is neither revoked nor
expired.

\myparagraph{Log Consistency}
After validation succeeded, the client saves the signed root for future
consistency checks. Then periodically, the client contacts a monitor to compare
the obtained root with the monitor's version.  If two roots with the same timestamps
are different, it means that the log misbehaved, which can be proved and reported
(e.g., to a software vendor).  To strengthen consistency checking, \name can be
enhanced by a system such as ARPKI~\cite{ARPKI}, or by
gossip protocols as proposed by Chuat et al.~\cite{gossip}.

%% file: deployment.tex
The deployability of a system like \name depends on many factors, such as the
incentives of the different parties to adopt the technology and the number of required
parties.
CT introduced two ways of providing proofs that a certificate is logged to clients
while preserving privacy~\cite{rfc6962,laurie2012certificatve}.
We describe these models in the context of
\name in the following two subsections. We also show that the deployment of \name is
challenging with one of the models introduced by CT, and the main reason for this
is that the ultimate goals of the two systems are different (CT tries to detect
misbehaving CAs, while a revocation system tries to avoid using invalid certificates).
However, we present new models including a browser-driven deployment that brings many
advantages, and a new lightweight realization of a log monitor. The presented deployment models
can also be used in conjunction.

\subsection{Server-Driven Deployment}\label{sec:server_deploy}
\begin{figure}[h!]
\centering
\includegraphics[width=0.55\columnwidth]{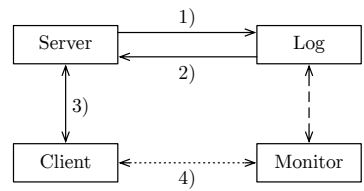}
\caption{Server-driven deployment model. Dotted and dashed lines represent
optional and periodic communications, respectively.}
\label{fig:server-deployment}
\end{figure}

In the first deployment scenario, depicted in
Fig.~\ref{fig:server-deployment}, servers are driving the process of proving
to their clients that their certificate is not revoked:

\begin{enumerate}
    \item The server contacts the log at regular intervals (at least every
        scheduling period) to obtain a fresh signed tree root and a fresh
        presence proof.
    \item The log returns the requested data.
    \item Every time a client connects to the server, this data, together with the
    certificate chain and the CC, is transmitted to the client (e.g., via an OCSP-stapling
    mechanism). 
    \item Clients can communicate with monitors to verify that they share a consistent
    and compatible vision of the log.
\end{enumerate}

This deployment model is ideal in terms of efficiency (because only the server
needs to periodically perform a few extra connections and the storage
requirements are low) and privacy (because the client does not need to contact a third party
to verify the validity of a server certificate).  However, this
model requires that servers are updated and this is not likely to happen
rapidly for all TLS servers on the Internet.

\subsection{ISP-Driven Deployment}\label{sec:isp_deploy}
As many servers are not updated regularly, the burden of contacting the log
to retrieve the revocation information could be put on clients, but there is
a privacy issue if clients do so directly. The documentation of CT~\cite{laurie2012certificatve}
mentions that clients could use a modified DNS resolver (provided by their
ISPs) as an intermediary to contact the log.
However, this model is problematic when it comes to revocation, since the
goal is no longer to simply detect attacks in an unspecified future, but to
instantly determine if a certificate can be considered valid.
Moreover, in a revocation system, such a connection would be required after
the certificate chain is received and before it is accepted (otherwise the client
does not know for which certificate chain a validity message should be
returned), which would increase latency and be prone to blocking attacks. For
these reasons, it would be challenging to adapt this model in \name.

\subsection{Browser-Driven Deployment}\label{sec:browser_deploy}
Since the ISP-driven deployment does not fit the requirements of \name,
and since we cannot assume that all servers would quickly be configured to provide 
fresh proofs to TLS connections (server-driven model), we
seek an alternative solution. We present a variant of browser-driven
deployment with the goal of providing users with the minimal
information required to ensure that no certificate (from the chain) is revoked.
To achieve this goal, we propose to extend a browser update mechanism
(mentioned in \S\ref{sec:pre:related}) that is already deployed,
namely CRLSets.

As in a browser-driven deployment clients are periodically provided with
revocation messages, it is crucial to minimize bandwidth and
storage overheads. In our deployment model, vendors employ the log as a source of
new revocations, and they push CRLSets that consist of identifiers (in our case
hashes) of all revoked and non-expired certificates with their corresponding
legitimacy periods (for CA certificates). Additionally, vendors are obliged to
log every CRLSet before it is propagated to the browsers, and are obliged to
propagate the CRLSet with a commitment (or presence proof) from the log,
indicating that the CRLSet is accepted by the log and will be visible in the near
future. We call this concept a \textit{Transparent CRL} (TCRL). On the log-side,
the TCRL is simply appended to the TimeTree. 
This deployment model does not provide properties as strong as the server-driven
deployment, but it allows to verify certificate validity and it enables the
audit of TCRLs.

\begin{figure}[h!]
\centering
\includegraphics[width=0.65\columnwidth]{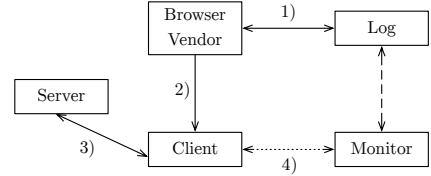}
\caption{Browser-driven deployment model. Dotted and dashed lines represent
optional and periodic communications, respectively.}
\label{fig:vendor-deployment}
\end{figure}

The connection establishment and certificate validation of this deployment model
are presented in Fig.~\ref{fig:vendor-deployment} and proceed as follows:
\begin{enumerate}
    \item Periodically, the browser vendor contacts a log to obtain the new
    revocations (note that a vendor can also act here as a monitor).
    \item The vendor prepares a software update creating a list of new
    revocations (TCRL). Then this TCRL is submitted to the log, which returns a commitment
    indicating that the TCRL will be appended to the TimeTree. Finally, with this commitment
    (or a presence proof), the vendor pushes the TCRL to the browsers.
    Browsers verify whether log proofs matched the TCRL and accept the update.
    \item During the TLS handshake, the client obtains a certificate chain along
    with the corresponding CC.\footnote{If a server deploys \name, then a proof
    is sent during the
    handshake, and the client validates the certificate chain as in
    \S\ref{sec:server_deploy}.}
    Then, with a locally-stored TCRL,
    the browser verifies whether all certificates from the chain have not been
    revoked. (As TCRLs provide complete revocation messages, clients
    can determine the legitimacy periods.) The browser continues with a verification
    similar to Algorithm~\ref{alg:validation}. 
    
    \item The browser can (optionally) contact a monitor, to
    verify that the local version of the TCRL (vendor's view) is consistent with
    the monitor's view. Note that this communication does not reveal any
    information on the domains that the browser has contacted.
\end{enumerate}

\subsection{Lightweight Monitoring}\label{sec:light_monitor}
Monitors are an integral part of many log-based schemes. They have the responsibility to
constantly monitor the logs to verify whether they behave correctly.
In previous proposals~\cite{AKI,ARPKI,laurie2012certificatve} monitors were
implemented as replicas of the logs that perform some extra checks on demand
(e.g., confirm that their view of the log is consistent with the root provided
by the client). Because of that design, the bandwidth and storage required to
operate a monitor are significant. In this section, we propose a novel
deployment model that allows to run a lightweight monitoring service.
This model could be used by network devices with security features or by power-users, for example.
Such a service can assist the clients in additional verification of a connection, and
the required features are:
\begin{itemize}
	\item confirm the root of the log,
	\item prove that the log is consistent (i.e., a version of the log is the extension of a previous one),
	\item prove that a given object is in the log.
\end{itemize}

Our first observation is that the large storage requirement of the log is
induced by the necessity of storing entire certificates (a single certificate
takes about 2 kB in PEM format). However, as \name clients are provided with
certificate chains and the corresponding information during TLS
connections, monitors need not store actual certificates but only the corresponding hashes.
This is sufficient to ensure that a certificate is indeed in the log and that the log is consistent.
In our proposal, a lightweight monitor is not directly equipped with the TimeTree's
leaves, but with their parent nodes (i.e., hashes) and with revocation messages.
Another observation is that certificates have a standardized maximum lifetime.
Therefore, after some time, the TimeTree will contain a continuous list of expired
certificates and there is no need to store the hashes of these certificates, unless
they are parts of non-expired chains.

\begin{figure}[h!]
\centering
  \begin{tikzpicture}[scale=.49,every tree node/.style={text width=0.8cm,font=\LARGE,anchor=base}]
    \Tree 
    [.\textit{root}
        [.$H_{abcdefr_0r_1}$
            [.\fbox{$H_\mathit{abcd}$}
        [.${H_{ab}}$
        [.$H_a$ $C_{a}$,\\$t_0$ ] [.$H_b$ $C_{b}$,\\$t_0$  ] ]
        [.${H_{cd}}$
        [.$H_c$ $C_{c}$,\\$t_0$ ] [.$H_d$ $C_d$,\\$t_0$ ] ] 
    ]
    [.${H_\mathit{efr_0r_1}}$
        [.\fbox{${H_{ef}}$}
        [.$H_{e}$ $C_e$,\\$t_0$ ] [.$H_f$ $C_f$,\\$t_0$  ] ]
        [.${H_\mathit{r_0r_1}}$
        [.\fbox{$H_{r_0}$} $r_0$,\\$t_0$ ] [.\fbox{$H_{r_1}$} $r_1$,\\$t_1$ ] ] 
    ]
 ]
  [.$H_\mathit{hijf'klmr_2}$
    [.$H_{hijf'}$
        [.${H_{hi}}$
        [.\fbox{${H_h}$} $C_h$,\\$t_2$ ] [.\fbox{$H_i$} $C_i$,\\$t_2$  ] ]
        [.${H_{jf'}}$
        [.\fbox{$H_j$} $C_j$,\\$t_2$ ] [.$H_{f'}$ \fbox{$R_{C_f}$}\\$t_2$ ] ] 
    ]
    [.${H_\mathit{klmr_2}}$
        [.$H_{kl}$
        [.\fbox{${H_k}$} $C_k$,\\$t_2$ ] [.\fbox{$H_l$} $C_l$,\\$t_2$  ] ]
        [.${H_{mr_2}}$
        [.\fbox{$H_m$} $C_m$,\\$t_2$ ] [.\fbox{$H_{r_2}$} $r_2$,\\$t_2$ ] ] 
    ]
  ]
 ]
\end{tikzpicture}
\caption{An example of a TimeTree, where all certificates before $t_1$ are
expired. Only nodes in boxes are stored by the lightweight monitor.}
\label{fig:browser_tree}
\end{figure}
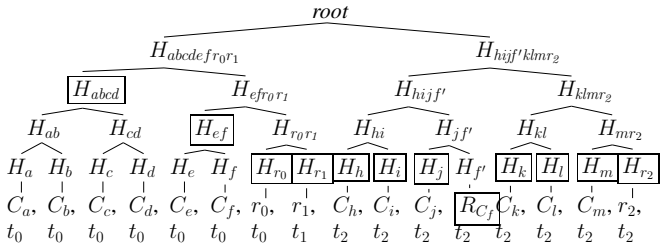

An example of our optimization is depicted in Fig.~\ref{fig:browser_tree}.
It shows the original TimeTree and the values that a monitor must provide.
In this case, a monitor must initially obtain from the log only the following:
\begin{equation}
\begin{aligned}
\label{eq:update}
    t_0 &: \{H_\mathit{abcd},H_{ef}, H_{r_0}\},\\
    t_1 &: \{H_{r_1}\}, \\
    t_2 &: \{H_h, H_i, H_j,R_{C_f}, H_k, H_l, H_m, H_{r_2}\},\\
        & \hspace{0.29cm}\{\textit{root}, t_2\}_{{k}_{\textit{log}}}.
\end{aligned}
\end{equation}
Then, periodically, a \textit{delta update} between the current TimeTree
and the monitor's local list is transferred. Every update is also accompanied with the
corresponding signed root (Eq.~(\ref{eq:root})). Such a design allows a monitor to
store a \textit{minimized} version of TimeTree and to:
\begin{itemize}
    \item check if every non-expired certificate of the chain is indeed present in the tree (e.g., on a client's query),
    \item check the revocations of certificates and determine legitimacy periods (e.g., on a client's query),
    \item build the TimeTree's root, and optionally compare it with other monitors to
verify that the view is consistent, 
    \item extend the tree with new hashes,
    \item verify the proofs received from the clients.
\end{itemize}

In this setting, a monitor is able to verify 
millions of certificates and needs to store only tens of megabytes, instead of
several gigabytes for a complete TimeTree. A detailed analysis of the required
resources
is presented in \S\ref{sec:eval:storage}. 
Moreover, such an optimization can be easily applied to other 
log-based approaches that employ hash trees.

%% file: analysis.tex
Our first claim is that \name provides authenticity, i.e., \textit{a non-capturing
adversary cannot create any legitimate revocation message}, as long as he cannot
forge a digital signature. An adversary with the private key of the
domain can revoke only the domain's certificate. However, by this action, an
attacker would reveal that the key is compromised, as the revocation must be
logged.

A more powerful adversary, able to capture a CA's private key, can revoke that
CA's certificate, and all its child certificates. We claim that \name provides
backward availability and timeliness, i.e., \emph{such an
adversary can misbehave only for a short time period} (e.g., by temporarily
introducing collateral damage or malicious certificates).  Specifically, that
time period is less than or equal to $T_d+T_a+T_s+T_p$, where $T_d$ is a
detection time, i.e., the duration between the moment when a misbehavior
(illegitimate revocation or certificate issuance) is logged and the moment when
the CA notices that misbehavior. $T_a$ denotes the audit delay, i.e., the time
during which the CA determines when the first misbehavior was logged, $T_s$ is
the scheduling period of the log (see \S\ref{sec:overview}).  For existing CT
logs a scheduling period (called in CT MMD) is set between 1-24 hours.  $T_p$
stands for the propagation time, i.e., the time it takes for a new change to be
propagated to clients. This time depends on a deployment model, however in all
presented models (see \S\ref{sec:deployment}) we may expect this time to be
bounded by a few hours.  Overall, we estimate that it is feasible to conduct the
entire process within several hours.

Let's consider the extreme case in which such an adversary compromises the root
CA's private key and revokes all child certificates with a revocation timestamp
close to the creation time of the CA. This would invalidate all the actions of
this CA. With \name, these revocation messages must be submitted to the log. The
log accepts them if they are signed with an authentic key.  These revocation
messages will be visible, at the latest, when the log is updated. After the
update, the malicious revocations are noticed by the CA, which, after an audit
procedure, can estimate when the breach happened, and can revoke its own
certificate with a revocation timestamp set to the breach time, using the
offline revocation key. Thereafter, in the next update of the log, all malicious
revocations will be invalidated, and this change will eventually be propagated
among clients (see \S\ref{sec:deployment}).  In general $T_a > T_d$, but when a
CA is revoked, or many revocations are submitted with a single key, the log
could inform the CA about these actions before the update. Such an information
would give a CA some time to take actions in order to completely eliminate the
collateral damage.  

As explained above, \name enables to remove collateral damage from the TLS PKI,
but with the assumption that the log is not malicious. We stress that the log
itself is only trusted to a certain extent, as it is constantly monitored and is
only supposed to: 1) be append-only, 2) accept object registrations, 3) return
cryptographic evidence about the content/consistency of the trees. Hence, the
log cannot revoke certificates by itself, as it requires a private key to sign
appropriate revocation messages. However, a misbehaving log can block requests
(by simply ignoring them), which is a more generic problem of all log-based
schemes.

The combination of a capturing adversary and a malicious log is especially
dangerous.  Consider the case (similar to the previous one) in which an
important CA is compromised and malicious revocation messages for child
certificates are issued and logged. Then, when the CA wants to revoke its own
certificate, the malicious log can just ignore the requests. As a consequence,
the malicious revocations will not be invalidated.  This attack is simple and
severe, but to succeed, an adversary must compromise the CA and the log at the
same time.

\name requires that all actions are signed and logged, \textit{making the parties
accountable}.  Revocations as well as certificates are \textit{transparent and
visible}, which makes \textit{split-world
attacks}~\cite{Mazieres:2002:BSF:571825.571840} detectable. Consider the case
where an adversary controls the log and captures a server's old revoked key.
Now, the adversary can produce a single fake presence proof, which states that a
given certificate is not revoked for example, and can launch a man-in-the-middle
attack on clients. Then, with such a proof, the adversary must provide the
corresponding signed root to the attacked client. The attack can succeed, as the
client trusts the log, but the attack is detectable if the client contacts a
monitor (or any other party) which has a different (legitimate) view of the log.
Such an attack is more difficult to conduct with the deployment scenario
sketched in \S\ref{sec:browser_deploy}, as revocations are stored in the
browsers.

\name \textit{preserves user privacy}. In all the presented deployment models
(\S\ref{sec:deployment}), clients receive complete revocation status either through
browser update or directly from the contacted server.  Clients do not contact
any third parties to ensure that a given certificate is valid.  Clients obtain
signed roots and extension proofs from the monitors, but this action also does
not reveal any information about websites visited.

%% file: implementation.tex
\subsection{Implementation}
In order to prove the feasibility of \name, we implemented the system in
Python~(2.7.6) and C++~(gcc-4.8.2), using the M2Crypto, libpki, and OpenSSL (1.0.1f)
cryptographic libraries. We modified libpki to add a
dedicated revocation key into the extension field of every CA X509v3
certificate~\cite{citeulike:6983793}. To minimize overheads we decided to use
the Ed25519 signature scheme, except for the standard keys of X509v3 certificates
where RSA-4096 was used instead. We used the SHA-256 hash function for
both certificates and the implementation of hash trees.

We wrote a complete log and TLS client that implements the validation logic from
\S\ref{sec:validation}. For the server side, we used Nginx, which
periodically requests a fresh presence proof and signed root from the log.
For every subsequent TLS Handshake, the server sends these values
(and the chain commitment) using \texttt{TLS Certificate Status
Request}~\cite{santesson2013x}, while the server's certificate is sent within a
standard \texttt{ServerHello} message. Such a configuration enables deployment of
\name without any changes to the TLS protocol. This setting is specific to the
deployment scenario presented in \S\ref{sec:server_deploy}.

\subsection{Performance}\label{sec:impl:perf}
With the setting presented above, we measured the efficiency of our
system by conducting a series of experiments. Every presence proof in our test
contained two revocation messages (pessimistic setting) and every certificate
chain contained three certificates. All results were obtained by
executing a given operation one thousand times on one Intel i5-3380M core @
2.90 GHz, on Ubuntu 14.04 with 16 GB of RAM. During one second,
the log was able to register 1907 certificate chains on average. For these
registrations, the log verified the chains and returned signed chain commitments. To add
10000 new certificate chains to the trees, and to update the trees, the log
needed on average 3.154 seconds. Our client's implementation conducted a complete
validation within 1.266 ms on average, where the pre-validation and proofs validations
take 0.405 ms and 0.370 ms, respectively. This computational overhead
should be unnoticed by users~\cite{journals/computer/ToliaAS06}.

%% file: eval.tex
We evaluate \name in terms of storage and bandwidth overheads and focus on the
server-driven deployment model, the browser-driven model, and the lightweight-monitor
proposal. For the server-driven deployment, the information required to
verify a certificate chain is obtained directly from the TLS Handshake. For the
lightweight monitor deployment, the monitor is provided with a delta update, as
in Eq.~(\ref{eq:update}), which allows the browser to reconstruct minimized
trees. In the browser-driven deployment~(TCRL), the browser receives
a delta update from the vendor as hashes of revoked certificates. Note that
these two variants provide different properties (see
\S\ref{sec:browser_deploy}). In our simulations, we assume that the
Ed25519~\cite{bernstein2012high} scheme is used as the signature scheme and that
the hash function produces an output of 20 bytes (this is a parameter, and
second pre-image resistance is the main property we rely on).

\subsection{Storage}\label{sec:eval:storage}
The Server-driven deployment does not require any storage on the client-side, and only
a small amount of storage on the server-side: a signed root (88 bytes), a chain commitment (96 bytes
for a chain of three certificates), and a presence proof (each node takes 20 bytes).
In the standard case, this overhead should be around 1~kB.

To estimate the storage overhead required for the presented deployment
variants, we used data available from one of CT's public
logs.\footnote{ \url{http://ct.googleapis.com/pilot}}  First, we
conservatively qualified certificates as valid considering their
\texttt{NotBefore} and \texttt{NotAfter} validity fields, and found that out of
the 7,427,474 certificates in the log, 3,938,656 were valid on 15 May 2015,
12:00:00 UTC (note that certificate chains can be added to the log only if the
root certificate is contained in a set of acceptable roots that the log
maintains). Then, we simulated storage overheads for the two deployment
variants, depending on the number of certificates and the fraction of revoked
certificates (this fraction in HTTPS was recently reported as 8\%~\cite{liu-2015-revocation}).

\begin{figure}[t!]
  \centering
  \includegraphics[width=\linewidth]{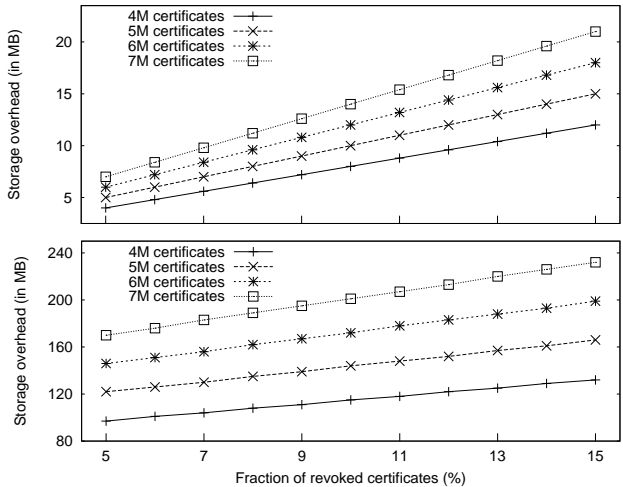}
  \caption{Storage overhead required by TCRL-enabled browser (top chart),
  and by a monitor with the minimized-trees variant (bottom chart).}
  \label{fig:storage}
\end{figure}

As shown in Fig.~\ref{fig:storage}, the results differ significantly depending
on the deployment variant.  With today's number of valid certificates and a 10\% revocation rate
(which is considered as high), a browser employing the TCRL mechanism needs 8 MB of storage,
while, for the same scenario, a lightweight monitor needs 115 MB,
whereas the log in such a setting stores about 8 GB.

\subsection{Bandwidth}\label{sec:eval:bw}
We also evaluated \name in terms of bandwidth required, using real-world traces.
Zhang et al.~\cite{Zhang:2014:ASC:2663716.2663758}, using data gathered by
Rapid7,\footnote{ \url{https://scans.io/study/sonar.ssl}} collected
information about certificates and the corresponding revocations. The certificates
were filtered to consider only valid ones from the Alexa Top 1 Million global
sites.\footnote{ \url{http://s3.amazonaws.com/alexa-static/top-1m.csv.zip}}
For these 628,692 certificates, the 1,386 corresponding CRLs were downloaded and processed.

The dataset we
used\footnote{ \url{https://ssl-research.ccs.neu.edu/dataset.html}} covers a
period from 30 October 2013 to 28 April 2014. This period is especially
interesting from our point of view as \textit{Heartbleed}---a critical
vulnerability in an OpenSSL extension---was publicly announced in April 2014.
Heartbleed allowed attackers to remotely read a server's protected memory
including sensitive information like private keys.
As a consequence, in mid-April 2014 we observed the highest
frequency of certificate re-issuance and revocation ever. This unique event and
its impact on the TLS ecosystem has been thoroughly
analyzed~\cite{Zhang:2014:ASC:2663716.2663758,durumeric2014matter}.

We evaluated the bandwidth required by \name during normal operations (i.e., a
few months before Heartbleed) and during what we will refer to as the
\textit{peak time} (i.e., right after Heartbleed was announced). For this test, we
used the above-mentioned dataset to extract all new certificate issuances and
revocations observed over the time period. We assumed that certificate chains
consist of three certificates, as this is, reportedly, the length of the vast majority
(about 98\%) of certificate chains~\cite{Holz:2011:SLT:2068816.2068856,Durumeric:2013:AHC:2504730.2504755}.
By fetching all entries from one of CT's public logs (as in \S\ref{sec:eval:storage}),
we determined that the average size of a single certificate is about 1966 bytes.
The setting of cryptographic primitives used here is the same as in the
previous test.

First, we estimated the total bandwidth required by the log to register all
issued certificates and revocations.
\begin{figure}[t!]
  \centering
  \includegraphics[width=\linewidth]{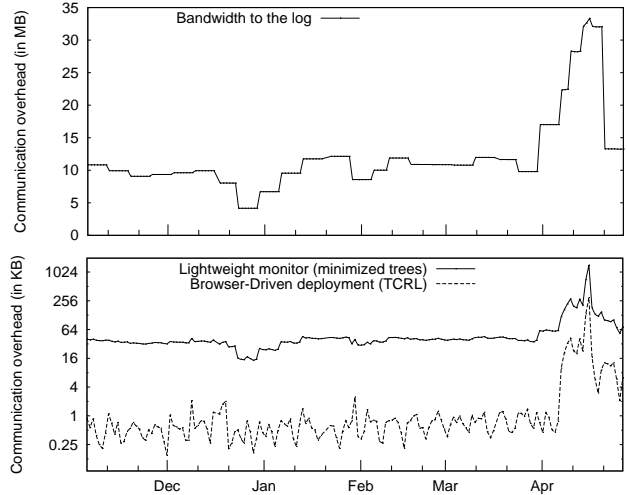}
  \caption{Bandwidth required by the log to receive certificate registrations
  and revocations (top chart), and by the browser to receive daily updates
  (bottom chart, note that the y-axis is in logarithmic scale).}
  \label{fig:bw}
\end{figure}
The results are presented in Fig.~\ref{fig:bw}. During the normal period
(between November 2013 and March 2014), the log receives 5--13 MB per day. At
peak time, the number of certificate issuances and revocations increases,
causing higher demands on bandwidth. However, even then, the maximum bandwidth
required is less than 35 MB per day.

Second, we estimated the bandwidth required for the daily update of a browser (TCRL) and a lightweight monitor (minimized trees). Fig.~\ref{fig:bw}
depicts the results for these two variants.

In a standard scenario, the daily update for the minimized trees variant is
15--40 kB, but with the increasing number of revocations caused by
Heartbleed, the required bandwidth increases as well. On 17 April 2014, it reaches
around 1.4 MB, which is the highest number observed. After this date, the
bandwidth required decreases rapidly. In a similar manner, for the deployment variant
using TCRL, the normal update is below 1 kB, while the update during the peak
reaches 300 kB at most.
We believe that such overhead is acceptable, but we expect that with a higher
revocation rate (which may occur in practice) browser vendors would reduce the
transfer cost through a more efficient encoding of TCRLs or by limiting the scope of TCRLs
(e.g., to EV certificates only---see CRLSets in \S\ref{sec:pre:related}).

In the server-driven deployment, for every TLS connection, a client is provided
with about 1 kB (see \S\ref{sec:eval:storage}) of additional data.

\subsection{Comparison}
We now summarize the above results and compare the different deployment models
of \name with competing revocation schemes.  The comparison encompasses storage
and bandwidth overhead on the client-side, as well as the potential latency
introduced by the revocation scheme to the TLS connection.
\begin{table}[b!]
\footnotesize
\centering
\renewcommand{\arraystretch}{1.1}
\caption{Comparison of revocation schemes.}
\label{tab:compar}
\begin{tabular}{@{~}lrrl@{~}}
\toprule
\textbf{Scheme} &\textbf{Storage} & \textbf{Bandwidth} & \textbf{Latency} \\
\cmidrule{1-4}
CRL         & 34 MB& 24 kB/conn.& increased \\
OCSP         & None & 0.5 kB/conn.& increased \\
OCSP Stapling& None & 0.5 kB/conn. & unaltered \\
CRLSet       & 0.2 MB & 0.12 kB/day   & unaltered \\
ECT/DTKI  & None & 1 kB/conn.  & increased\\
AKI/ARPKI  & None & 0.5 kB/conn.  & unaltered \\
\name (srv-driven, \S\ref{sec:server_deploy})/PoliCert & None & 1 kB/conn.  & unaltered \\
\name (browser-driven, \S\ref{sec:browser_deploy}) & 6.4 MB & 0.7 kB/day & unaltered \\
\name (light. monitor, \S\ref{sec:light_monitor}) & 108 MB & 39 kB/day & unaltered \\
\bottomrule
\end{tabular}
\end{table}
The results are presented in Table~\ref{tab:compar}.  Depending on the scheme,
the revocation information can be passed through an update (e.g., daily) or
during every TLS Handshake (per connection), which is described in the Bandwidth
column.  For \name and other log-based approaches we show the storage
required for a revocation rate of 8\% and four million active certificates (see
\S\ref{sec:eval:storage}). The bandwidth required by \name is given
as the median value observed in \S\ref{sec:eval:bw}, while for CRLSets we used the
dataset provided by Liu et al., and for CRLs we used a dataset provided by
ISC~\cite{isc_crl}.

Besides efficiency, the schemes compared here differ significantly in the properties
they offer (see \S\ref{sec:pre:related} and \S\ref{sec:deployment}).

\subsection{Case Study}

GoDaddy is currently one of the largest issuers
of TLS certificates~\cite{Arnbak:2014:SCH:2668152.2673311}.
We take the ``Go Daddy Secure Certification Authority'' certificate
(serial number \texttt{07969287}) as an example in a case study on how effective \name could
be in practice. By analyzing the content of Google's pilot CT log, we found
139,086 valid certificates (on 19 November 2015) signed by the aforementioned intermediate CA.
The oldest of these certificates (as indicated by the \texttt{NotBefore} field)
was issued on 29 January 2007, which means that a single private key was used to
sign about 43.25 certificates per day on average, during more than 8 years.
If that key was compromised and the corresponding certificate revoked with current
methods, thousands of websites would be affected. With \name, only a small number of
certificates would be revoked (provided that the detection process is reasonably
fast). For instance, if a misbehavior was detected after one week, only about 300
certificates would have to be revoked and re-issued, which constitutes only about
0.2\% of all certificates issued with this key.

%% file: discussion.tex
The effectiveness and security of our system depend on the length of update periods,
which introduces an obvious trade-off between the log's performance and the size
of the attack window. We believe that a delay of a few hours between
log updates is a good compromise.

One remaining challenge, and a potential subject for future work, is the multi-log
scenario, which is challenging as synchronization between
the logs would be necessary. 
One interesting approach to make the multi-log scenario scalable, is to introduce
domain-driven security policies~\cite{PoliCert} that would allow domains to specify
which logs they trust. Then, all certificate registrations and revocations could
be submitted only to these logs. Another interesting aspect that could be investigated
relates to the question of how \name can be extended to other trust models, log systems,
and their applications~\cite{melara2014,Fahl:2014:HNS:2660267.2660311}. In particular,
\name could be combined with ARPKI~\cite{ARPKI}, for example, to provide additional
security properties (such as ``connection integrity'').
We also plan to conduct a formal analysis of \name.

An open problem, that all new log-based approaches face, is to find an optimal
deployment model and an incremental deployment plan. \name can benefit from the
previous works~\cite{matsumoto2015deployment,Bates:2014:SSC:2660267.2660338}, but we plan to
investigate and analyze the proposed deployment models in depth.
An advantage of \name is that it can be easily built on the top of CT, which
currently is being deployed.

We believe that the revocation policy employed by \name fits the current TLS ecosystem
and reflects the power of PKI actors and the connections between them.
However, we envision that this policy could be optimized and standardized by
organizations and consortia such as the CAB Forum~\cite{cabf}.

%% file: conclusions.tex
The current certificate revocation systems suffer from many drawbacks such as large
attack windows, privacy issues, and configuration dependencies. 
In this paper, we redesigned the current TLS revocation system and presented \name,
which resolves several problems that we identified.
The most important advantage of \name is that it is the first system (to the best of our knowledge)
to solve the too-big-to-be-revoked problem of the current PKI.
It also enhances transparency and introduces a novel revocation policy that
reflects the actual interactions within the TLS ecosystem.
Only a few changes are required to deploy \name with the current infrastructure.
Moreover, the evaluation and performance results of our implementation indicate that \name
is viable for use in practice.